\documentclass[useAMS,usenatbib]{mn2e}
\usepackage[dvips]{graphicx}
\usepackage{amssymb}
\usepackage{txfonts}
\usepackage{float}

\title[The spectrum of NGC\,55\,C1\_31]{First VLT/X-shooter spectroscopy of early-type stars outside the Local Group\thanks{Based on observations made with ESO Telescopes at the Paranal Observatory under program 60.A-9419(A)}}

\author[O.E. Hartoog et al.]{O. E. Hartoog$^{1}$ \thanks{O.E.Hartoog@uva.nl}, H. Sana$^{1}$, A. de Koter$^{1,2}$ , L. Kaper$^{1}$ \\
$^{1}$ Astronomical Institute ``Anton Pannekoek", University of Amsterdam, Science Park 904, 1098 XH Amsterdam, The Netherlands\\
$^{2}$ Astronomical Institute, Utrecht University, Princetonplein 5, 3584 CC Utrecht, The Netherlands\\
}
\begin{document}

\pagerange{\pageref{firstpage}--\pageref{lastpage}} \pubyear{2012}

\maketitle

\label{firstpage}

\begin{abstract}
As part of the VLT/X-shooter science verification, we obtained the first optical medium-resolution spectrum of a previously identified bright O-type object in NGC\,55, an LMC-like galaxy at a distance of $\sim$2.0\,Mpc. Based on the stellar and nebular spectrum, we investigate the nature and evolutionary status of the central object(s) and its influence on the surrounding interstellar medium. We conclude that the source, NGC\,55\,C1\_31, is a composite object, likely a stellar cluster, which contains one or several hot ($T_\mathrm{eff}\simeq$ 50\,000~K) WN stars with a high mass-loss rate ($\sim$$3\times 10^{-5} \mathrm{~M}_{\odot} \mathrm{~yr}^{-1}$) and a helium-rich composition ($N_{\mathrm{He}}/N_{\mathrm{H}}=0.8$). The visual flux is dominated by OB-type (super)giant stars with $T_\mathrm{eff}\lesssim$ 35\,000~K, solar helium abundance ($N_{\mathrm{He}}/N_{\mathrm{H}}=0.1$), and mass-loss rate $\sim$$2 \times10^{-6}\mathrm{~ M}_{\odot}$~yr$^{- 1}$. 
The surrounding  H\,{\sc ii} region has an electron density $n_\mathrm{e} \leq 10^2$~cm$^{-3}$ and an electron temperature $T$(O\,{\sc iii}) $ \simeq 11\,500 \pm600$~K. The oxygen abundance of this region is $[\mathrm{O}/\mathrm{H}]=8.18\pm 0.03$ which corresponds to $Z=0.31 \pm 0.04\,\mathrm{ Z}_{\odot}$. We observed no significant gradients in $T$(O\,{\sc iii}), $n_\mathrm{e}$ or $[\mathrm{O}/\mathrm{H}]$  on a scale of 73~pc extending in four directions from the ionising source.  The properties of the  H\,{\sc ii} region can be reproduced by a CLOUDY model which uses the central cluster as ionising source, thus providing a self-consistent interpretation of the data.
We also report on the serendipitous discovery of He\,{\sc ii} nebular emission associated with the nearby source NGC\,55 C2\_35, a feature usually associated with strong X-ray sources.

\end{abstract}


\section{Introduction}
\label{sec:intro}

The most luminous stars in low metallicity galaxies are of special interest. These may have masses exceeding $200~\textrm{M}_\odot$, defining the upper mass limit of stars. Until recently, such massive objects were only found in the cores of young massive clusters \citep{crowther2010}, but the first such object has now been found in apparent isolation \citep{bestenlehner2011}. This motivates a search for very massive stars in Local Group dwarf galaxies, or even more distant systems, where current instrumentation does not yet allow stellar clusters to be spatially resolved. Depending on mass and metallicity, the mass-loss rates of the brightest stars may be so high that their winds become optically thick, resulting in hydrogen-rich WN spectra \citep{dekoter1997}. Such targets provide important tests for the theory of line-driven winds \citep{grafener2011,vink2011}. Massive stars up to about $80~\textrm{M}_\odot$ in metal poor environments receive special attention as well, as those that have a rapidly rotating core at the end of their lives may produce broad line Type Ic supernovae or hypernovae, which are perhaps connected to long duration gamma-ray bursts \citep{moriya2010}.\\

In this context, massive stars in the Magellanic Clouds have been extensively studied \citep[see e.g.][]{evans2004,evans2011}. 
Extending such studies to more distant galaxies requires sensitive spectrographs mounted at the largest telescopes and, so far, has only be attempted at low resolution \citep{bresolin2006,bresolin2007,castro2008}. Such low resolution studies can be hampered by the presence of nebular emission, as due to their strong UV flux, massive stars ionise the ambient environment creating H\,{\sc ii} regions. Through observations at higher spectral resolution, nebular emission, rather than being a complicating factor, may help to further constrain physical properties of the ionising source \citep[see e.g.][]{kudritzki1990}.\\

In this paper we take the first step towards quantitative spectroscopy of massive stars outside the Local Group. We present the first medium-resolution spectrum of a luminous early-type source in NGC\,55 ($\sim$2.0\,Mpc) and its surrounding region, obtained with the new X-shooter spectrograph at the ESO {\it Very Large Telescope} (VLT). The medium spectral resolution ($R \sim 6000$) and unique spectral coverage of X-shooter allow for a detailed analysis of the stellar spectrum, and, importantly, for an improved correction of the nebular emission lines that can be distinguished from the underlying stellar spectrum.

\subsection*{NGC~55}

\begin{figure}
\includegraphics[width=8.5cm]{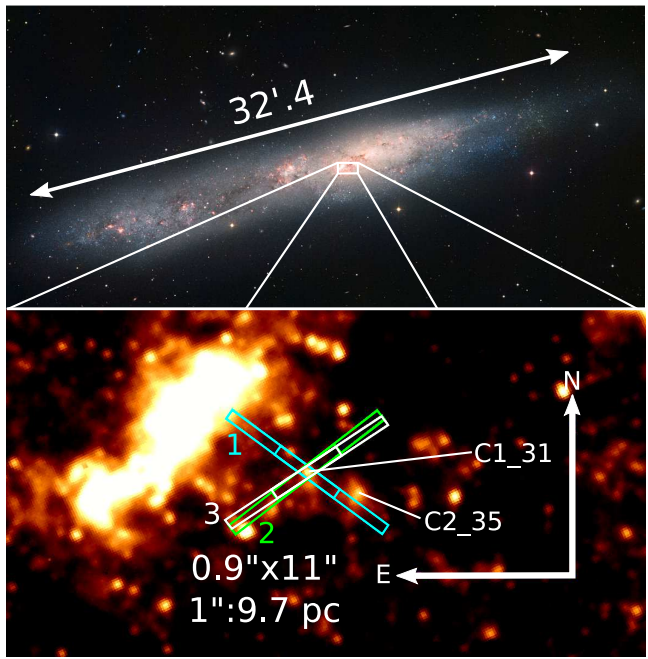}
\caption{$B-V-\mathrm{H}\alpha$ image of the host galaxy NGC\,55 (MPG/ESO, Wide Field Imager) with a zoomed in region (FORS H$\alpha$ image) that indicates the location of our source. The rectangles show the projections of the VIS entrance slit ($0.9''\times11''$) at the different observing epochs: (1) is 13/08/09, (2) is 27/09/09 and (3) is 30/09/09. C1\_31 is located where the slit projections cross. C2\_35 is another massive star candidate, of which we see the surrounding nebular emission in our slit (see Section~\ref{sec:C235}).}
\label{fig:regionslit}
\end{figure}

NGC\,55 (see Fig.~\ref{fig:regionslit}) is located in the foreground of the Sculptor Group, at a distance of approximately 2.0~Mpc \citep[][and refererences therein]{gieren2008}.  Though it is difficult to determine its morphological type due to its high inclination of $\sim$$80^{\circ}$ \citep{hummel1986}, the galaxy is likely of type SB(s)m: a barred spiral with an irregular appearance, very similar to the Large Magellanic Cloud \citep{devaucouleurs1972}. Metallicity measurements of NGC\,55 show a range of values between $0.23$ and $0.7\,\mathrm{ Z}_{\odot}$, all determined by analysis of forbidden oxygen line emission \citep[see e.g.][]{webster1983,stasinska1986,zaritsky1994,tuelmann2003}.\\

The blue massive star population of NGC\,55 has been studied in the context of the Araucaria project \citep{gieren2005}. As part of this project, \citet{castro2008} have presented low resolution ($R=780$) optical ($390-490$ nm) VLT/FORS2 spectra of approximately 200 blue massive stars in NGC\,55. In search for the most massive star in this galaxy, we selected NGC\,55\,C1\_31 (for the remainder of this paper C1\_31, RA 0:15:00.01, DEC -39:12:41.39, indicated in Fig.~\ref{fig:regionslit}) because of its brightness and its classification as an early O-type supergiant. \\

In the following section we describe the observations and data reduction for both the stellar spectrum (\S \ref{sec:redphotspec}) and the nebular spectra as a function of location along the slit (\S \ref{sec:rednebspec}). In Section~\ref{sec:photospec}, we constrain the overall properties of the central source by comparing the observed hydrogen and helium line profiles of C1\_31 with simulated profiles. In Section~\ref{sec:surrh2} we analyse the nebular spectra. In Section~\ref{sec:modelhii} we explore the effect of luminosity and temperature of the ionizing source on the properties of the surrounding nebula, resulting in a consistent picture on both the properties of the central ionizing source and the surrounding nebula (\S \ref{sec:consistentpic}). We summarise the main conclusions in Section~\ref{sec:conclusion}.

\section{Observations and Data Reduction}
\label{sec:obsanddatared}

	\begin{table}
	\caption{Overview of the observations. Date and mid-exposure time of the observations; Seeing measured from $R$ band acquisition images; Angular distance from the source to the Moon and fraction of lunar illumination (FLI); Exposure time per arm; Position angle (North to East) of the slit on the sky (along the parallactic angle at the time of the observations).\label{tab:observations}}
	\begin{tabular}{ cccccc }
\hline

&	Date			&Seeing			& Moon dist. 		&Exp. T.	& Pos. Ang.\\
&	Time (UT)		&	 ($R$-band) 	& FLI			& (s)			& 		\\
	\hline
1&	13/08/09 		& $0.69-0.88''$		& $69^{\circ}$			& $2 \times 900$	& $53.4^{\circ}$\\
&	08:52		&				& 56\%				& 			&\\	
2&	27/09/09		& $0.87-0.93''$		& $66^{\circ}$			& $4 \times 900$	&$-52.3^{\circ}$\\
&	03:43		&				& 60\%				& 			&\\
3&	30/09/09 		& $0.96-1.50''$		& $45^{\circ}$			& $4 \times 900$	&$-56.4^{\circ}$\\
&	03:26		&				& 85\%				& 			&\\
	\hline

	\end{tabular}
	\end{table}
	
	\begin{table}
	\centering
	\caption{Overview of the X-shooter instrument properties. Instrument arm; Wavelength range; Projected slit size; Measured resolving power $R=\lambda/\Delta\lambda$; Resolving power according to the X-Shooter User Manual.\label{tab:arms}}
	\begin{tabular}{ ccccc }

\hline
	Arm	&	Range (nm)	&Slit dimensions	& $R$			& $R_\mathrm{th}$ \\
	\hline
	UVB	& $300-560$		& $0.8''\times 11''$	&$6268\pm179$	& 6200	\\
	VIS	& $550-1020$		& $0.9''\times 11''$	&$7778\pm264$	& 8800	\\
	NIR	& $1020-2480$		& $0.6''\times 11''$	&$7650\pm258$	& 8100	\\
	\hline
	\end{tabular}
	\end{table}

The observations of C1\_31 were obtained as part of X-shooter Science Verification (SV) Runs 1 and 2. Spread over three nights, the total exposure time is 2.5h. The observations were carried out in nodding mode using a nod throw of $5''$. We refer to Table~\ref{tab:observations} for the details of the observations and the observing conditions. \\

Light that enters X-shooter is split in three arms using dichroics: UV-Blue (UVB), VISual (VIS) and Near-IR (NIR). Each instrument arm is a fixed format cross-dispersed  \'echelle spectrograph \citep{dodorico2006,vernet2011}. Table~\ref{tab:arms} gives for each arm the wavelength range, the projected dimensions of the slit and the resolving power. The data are reduced with the X-shooter pipeline version 1.2.2 \citep{modigliani2010,goldoni2011}. Although our source is observed in nodding mode, we have reduced the UVB and VIS science frames separately, using the staring mode reduction recipe. We follow the full cascade of X-shooter pipeline steps ('physical model mode'), up to obtaining two-dimensional (2D) straightened spectra, without sky subtraction. See Section~\ref{sec:redphotspec} for the steps through which we obtain the stellar spectrum of C1\_31, and Section~\ref{sec:rednebspec} for the extraction procedure of the nebular spectra along the spatial direction of the slit.\

\subsection{The sky-corrected stellar spectrum}
\label{sec:redphotspec}
\subsubsection*{UVB and VIS arms}
One-dimensional object and sky spectra are extracted from the 2D spectra. The sky spectrum is extracted from regions with the lowest continuum and nebular line emission. In the nights with four consecutive exposures, cosmic rays have been removed by taking the median value of each pixel in the four exposures. The sky spectra are subtracted from the object spectra, thereby correcting for the contamination by moonlight as well. After applying the barycentric correction, the sky-corrected object spectra of the three nights have been combined. Table~\ref{tab:snr} gives the signal-to-noise ratio (SNR) per resolution element of the result of each night and of the combined spectrum.  For the stellar line analysis we normalised the spectrum. 
Independently, we also calibrated the flux of the UVB and VIS spectra with standard star BD$+17^{\circ}\,4708$ (sdF8) taken on 27/09/09, since there were no appropriate flux standard observations for each individual night. We did not correct the spectrum for slit losses.
\subsubsection*{NIR arm}
We reduced the NIR spectra following the pipeline cascade for nodding mode up to extracted 1D spectra, and we combined the three nights. A telluric standard star (HD\,4670, B9 V) is used to correct for the telluric absorption features in the combined NIR spectrum, and to calibrate the relative flux. The NIR spectrum is scaled to match the absolutely calibrated VIS spectrum. The SNR of the combined NIR spectrum of all three nights is $\sim$3 in the $J$ band, and $<1$ in the $H$ and $K$ bands. This is too low to detect stellar or nebular features in the spectrum; however, the level of the continuum can be retrieved by binning the flux in the atmospheric bands. The result is shown in Fig.~\ref{fig:fullspec_dered}. 

\begin{table}
\caption{Signal-to-noise ratio (SNR) per resolution element of different parts of the UVB and VIS spectra per night, and of the final combined spectra.\label{tab:snr}}
		\begin{tabular}{ l @{ }c@{ } r r r  r }
		\multicolumn{6}{l}{}\\
		\multicolumn{6}{l}{\textit{UVB arm}}\\
						&&	\multicolumn{3}{c}{SNR [range (nm)]}& \\
		Night			&& [424:428] &	[460:465]	&	[505:510]	& \\
		\hline
		13/08/09	&&	16.9&	21.3	&	21.4	&	\\
		27/09/09	&&	14.0&	21.2	&	13.7	&\\
		30/09/09	&&	7.2&	9.8		&	9.8	&\\
		\hline
		combined		&&	21.2	&	29.3	&	27.4	&		\\
		\multicolumn{6}{l}{}\\
		\multicolumn{6}{l}{\textit{VIS arm}}\\
				&&	\multicolumn{3}{c}{SNR [range (nm)]}&\\
		Night	&&	[604:609]&	[675:680]&	[811:816]& [975:978] 	\\
		\hline
13/08/09	&&	12.1&	15.5&	22.0&	6.3	\\
27/09/09	&&	10.3&	12.0&	19.8&	5.2	\\
30/09/09	&&	6.2	&	9.8	&	13.3&	4.8	\\
		\hline
		combined 		&&	16.2	&	20.9	&	30.0	& 8.7		\\
		\end{tabular}
\end{table}

\subsection{The nebular emission spectra}
\label{sec:rednebspec}
The 2D spectra of each night are combined using the median value. We flux calibrate the combined 2D spectra with the same photometric standard star as we used for the object spectrum. Following the trace of C1\_31, we extract 20 sub-apertures from each night's combined 2D spectrum, resulting in a set of 1D nebular spectra that are spatially separated by $0.75''$ each\footnote{The sub-aperture size of $0.75''$ is just below the average FWHM of the seeing.}. Independently, we apply this combining and sub-aperture extraction procedure as well to the non-flat-fielded 2D spectra, now using the sum. This allows us to determine the number of photons $N$, and thus the photon noise error $\sqrt{N}$ for every emission line in the nebular spectra. These errors are propagated in the values of the nebular properties described in Section~\ref{sec:surrh2}. \\

\section{Analysis: The C1\_31 stellar spectrum}
\label{sec:photospec}

\begin{figure*}
\includegraphics[width=17cm]{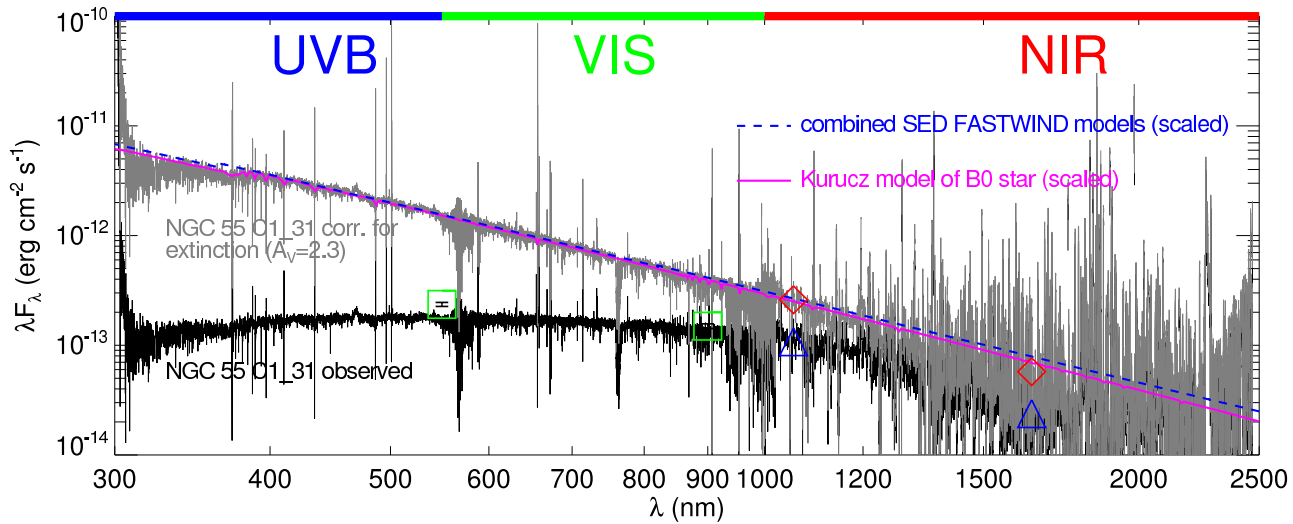}

\caption{Flux-calibrated spectrum of NGC55\,C1\_31 (bottom), and the same spectrum (top) corrected for extinction with $R_V=3.24$ and $A_{V,\mathrm{star}}=2.3$ (see text). The dashed line shows the FASTWIND spectral energy distribution for the best fitting combination of models (see Section~\ref{sec:consistentpic}); the solid line is a Kurucz model of a B0 star \citep{kurucz1979,kurucz1993}. The ranges of X-shooters instrument arms are indicated at the top of the graph. The open squares point to the updated $V$ and $I$ magnitudes (priv. comm. Pietrzy\'nski). For clarity, the NIR spectrum is shown here with a smoothing of 7 points. The triangles show the integrated values of the observed flux in the $J$ and $H$ bands; the diamonds show their extinction corrected values.}
\label{fig:fullspec_dered}
\end{figure*}

Fig.~\ref{fig:fullspec_dered} shows the combined flux-calibrated stellar spectrum of C1\_31. The extinction corrected flux should show a Rayleigh-Jeans wavelength dependence ($F_{\lambda} \propto \lambda^{-4}$), because we expect an early-type star based on the classification in  \citet{castro2008}. To lift the observed spectrum to the slope of the scaled Kurucz model of a B0 star, we need to de-redden our spectrum with $A_{V,\mathrm{star}}=2.3\pm0.1$, adopting $R_V=3.24$ \citep{gieren2008}. We apply the parametrized extinction law by \citet{cardelli1989}. As a check, we do the same exercise by varying both $R_V$ and $A_{V,\mathrm{star}}$. With $3.0 \lesssim R_V \lesssim 3.5$, the spectrum can be de-reddened to the intrinsic $F_{\lambda} \propto \lambda^{-4}$ slope with an $A_{V,\mathrm{star}}= 2.3\pm0.1$. With $R_V$ outside this range, the de-reddened spectrum does not match the models for any value of $A_{V,\mathrm{star}}$.\\

\citet{castro2008} report magnitudes $V=18.523$ and $I=19.239$ for NGC55\,C1\,31, which were obtained as part of the Araucaria Cepheid search project \citep{pietrzynski2006}. The $I$ magnitude is in good agreement with our flux-calibrated spectrum; $V$ is not. The reported color $V-I=-0.716$ is even bluer than a theoretical Rayleigh-Jeans tail, suggesting a problem with the photometry\footnote{This has been confirmed by the authors. Corrected photometric values are $V=19.87\pm0.05$, $I=19.25\pm0.05$ (priv. comm. Pietrzy\'nski), in agreement with our findings.}. From our flux-calibrated spectrum we  obtain $V\simeq20.1\pm0.1$. Using $A_{V,\mathrm{star}}=2.3$, and $d=2.0$\,Mpc, we derive an absolute magnitude of $M_V=-8.7\pm0.4$ for the source. \\

\par
\begin{figure}
\includegraphics[width=8.5cm]{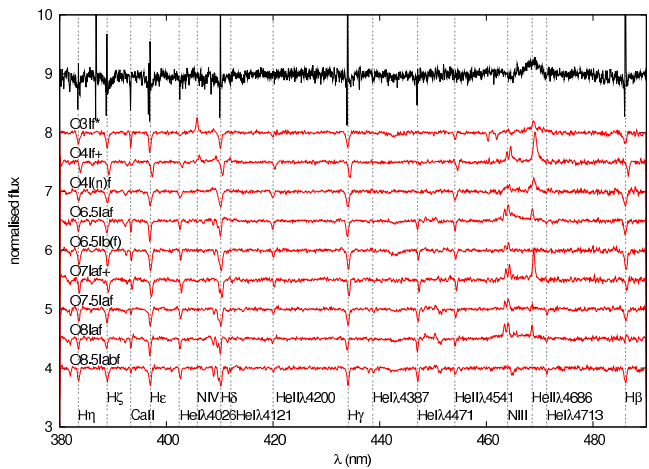}
\caption{The normalised sky-corrected object spectrum (top) between 380 and 490\,nm compared with the spectra of standard O supergiants from O3 to O8.5 \citep{walborn1990}. Note the very broad He\,{\sc ii}\,$\lambda$4686 emission.}
\label{fig:compare_obj}
\end{figure}

 Fig.~\ref{fig:compare_obj} compares the observed UVB spectrum of C1\_31 between 380 and 490 nm with that of spectral standard stars \citep{walborn1990}. In the C1\_31 spectrum the Balmer and He\,{\sc i} lines show artifacts of the nebular emission correction. However, most of the line wings are left unaffected thanks to the relatively high spectral resolution of X-shooter. Based on the non-detection of He\,{\sc ii}\,$\lambda$4541 and the presence of He\,{\sc i}\,$\lambda$4471 in the C1\_31 spectrum one can not classify this source as an early O-star. By adding artificial noise to the standard star spectra, matching the SNR of our observations in this range, we estimate that the He\,{\sc ii}\,$\lambda$4541 line of a star with spectral type later than O7.5 will not be detectable, thus suggesting a late spectral subtype. In Section~\ref{sec:modeltemp} we will use stellar atmosphere models to constrain the effective temperature more quantitatively. Supergiants may have strong emission in  He\,{\sc ii}\,$\lambda$4686 due to their stellar wind, but none of the standard stars show a feature as broad as that in our observations. Broad He\,{\sc ii}\,$\lambda$4686 emission lines are the strongest features in Wolf-Rayet stars of type WN. This line profile will be analysed in more detail in Section~\ref{sec:heii4686}. 

When we compare the X-shooter spectrum to the FORS2 spectrum of C1\_31 in \citet{castro2008}, we conclude that the general appearance is similar, including the shape of He\,{\sc ii}\,$\lambda$4686. \citeauthor{castro2008} observed He\,{\sc ii}\,$\lambda$4200 weakly in emission as well; this we can not confirm. The FORS2 observations will have suffered from similar nebular contamination in the Balmer lines as well as in some He\,{\sc i} lines, which may have hampered earlier classification, but correcting for this is even more difficult at lower spectral resolution.

\subsection{Stellar line profile modelling}
\label{sec:lineprofilemodel}

	\begin{table*}
	\caption{Parameters used for the described FASTWIND models, which are a subset of a larger grid. $R_*$, $\log g$ and $v_{\infty}=2.6 \times v_{\mathrm{esc}}$ depend on the other variables. The first four models have the same $M$, $L_*$ and $\dot{M}$, but different $T_{\mathrm{eff}}$ causing the remaining parameters to be modified. The next three models are variations on MOD31, but with different values for $\dot{M}$. MOD69 is like MOD31, but with an artificially high wind acceleration parameter $\beta$. The last two models MOD89 and MOD90 provide the best fitting multiple model configuration (in a visual flux ratio $\sim4:1$). The bolometric luminosities $L_*$ are chosen such that they reproduce the measured $M_V$ with ten times MOD89 and two times MOD90. \label{tab:modparam}}
	\begin{tabular}{ ccccccccccc}
	\hline
	ID 	&$M$	 & $T_{\mathrm{eff}}$	&$L_*$		&$\log g$			& $R_*$& $\dot{M}$& $v_{\infty}$ & $\beta$ & $N_{\mathrm{He}}/N_{\mathrm{H}}$	&$Z$\\
		&$(\mathrm{M}_{\odot})$	& (K)		&$(\mathrm{L}_{\odot})$& (cm s$^{-2}$)	&$(\mathrm{R}_{\odot})$	& $(\mathrm{M}_{\odot}\mathrm{~yr}^{-1})$	& (km~s$^{-1}$)	&	& & ($\mathrm{Z}_{\odot}$)\\
\hline
MOD39	&	40	&	27\,500	&	$10^{	5.6	}$	&	2.99	&	33.69	&	$6.00\times 10^{-6}$	&	1750.33	&	1.0	&	0.1	& 0.3\\
MOD31	&	40	&	30\,000	&	$10^{	5.6	}$	&	3.30	&	23.40	&	$6.00\times 10^{-6}$	&	2100.40	&	1.0	&	0.1	& 0.3\\
MOD32	&	40	&	32\,500	&	$10^{	5.6	}$	&	3.44	&	29.94	&	$6.00\times 10^{-6}$	&	2275.43	&	1.0	&	0.1	& 0.3\\
MOD33	&	40	&	35\,000	&	$10^{	5.6	}$	&	3.57	&	17.19	&	$6.00\times 10^{-6}$	&	2450.47	&	1.0	&	0.1	& 0.3\\
\hline
MOD10	&	40	&	30\,000	&	$10^{	5.6	}$	&	3.30	&	23.40	&	$1.00\times 10^{-6}$	&	2100.40	&	1.0	&	0.1	& 0.3\\
MOD11	&	40	&	30\,000	&	$10^{	5.6	}$	&	3.30	&	23.40	&	$3.00\times 10^{-6}$	&	2100.40	&	1.0	&	0.1	& 0.3\\
MOD12	&	40	&	30\,000	&	$10^{	5.6	}$	&	3.30	&	23.40	&	$1.00\times 10^{-5}$	&	2100.40	&	1.0	&	0.1	& 0.3\\
\hline
MOD69	&	40	&	30\,000	&	$10^{	5.6	}$	&	3.30	&	23.40	&	$6.00\times 10^{-6}$	&	2100.40	&	3.0	&	0.1	& 0.3\\
\hline
MOD89	&	30	&	30\,000	&	$10^{	5.24	}$	&	3.54	&	15.46	&	$2.00\times 10^{-6}$	&	2237.86	&	1.0	&	0.1	& 0.3\\
MOD90	&	80	&	50\,000	&	$10^{	5.83	}$	&	4.26	&	10.98	&	$3.00\times 10^{-5}$	&	4336.76	&	1.0	&	0.8	& 0.3\\
\hline
	\end{tabular}
	\end{table*}
	
Although we are going to make clear in this paper that C1\_31 is very likely a composite source, we first approach the spectrum as if it is produced by a single star. The derived physical parameters therefore represent averages of the flux-weighted components contributing to the spectrum. We note, though, that the integrated light from clusters  - in particular the hydrogen ionising radiation -  is often dominated by only a few of the most massive components.\\

The profiles of spectral lines are responsive to various stellar parameters such as effective temperature, mass-loss rate, surface gravity, chemical abundances and rotation speed. We have used FASTWIND \citep{puls2005} to model stellar atmospheres and to compute profiles of spectral lines. FASTWIND calculates non-LTE line blanketed stellar atmospheres and is suited to model stars with strong winds. 
We first tried to apply a genetic fitting algorithm with FASTWIND models \citep[see][]{mokiem2005} to the observed spectrum, in order to fit a large number of parameters at the same time. This did not result in well constrained parameters, because the shape and width of He\,{\sc ii}\,$\lambda$4686 could not be fitted at the same time as the other lines. \\

Instead we constructed a grid of FASTWIND models, and constrained the parameters by comparing the observed profiles with the models. The main grid varies effective temperature $T_{\mathrm{eff}}=27\,500 - 35\,000$\,K with steps of $2\,500$\,K; and covers values for the mass-loss rate $\dot{M}=1,3,6$ and $10 \times 10^{-6}\mathrm{~M}_{\odot}\mathrm{~yr}^{-1}$ and luminosity $\log(L_*/L_{\odot})=5.4,5.6$ and $5.8$. All models have a mass $M=40\mathrm{~M}_{\odot}$, wind acceleration parameter $\beta=1.0$, and helium to hydrogen number density $N_{\mathrm{He}}/N_{\mathrm{H}}=0.1$. Radius $R_*$ and surface gravity $g$ are fixed by the other parameters. The terminal wind velocity $v_{\infty}$ is assumed to be 2.6 times the surface escape velocity \citep[see][]{lamers1995}.
In Table~\ref{tab:modparam} we show the parameter values of a subset of the grid, i.e. the models discussed in more detail in this paper.  
Unless stated otherwise, the synthesized line profiles are produced using a microturbulent velocity $v_\mathrm{turb}=10$\,km\,s$^{-1}$ and projected rotational velocity $v_{\mathrm{rot}} \sin({i})=150$\,km\,s$^{-1}$, with $i$ the inclination of the stellar rotation axis relative to the plane of the sky. An instrumental profile matching the resolution of X-shooter is applied as well.\\

We first investigate the overall impact of the effective temperature (\S\ref{sec:modeltemp}), rotational velocity (\S\ref{sec:vrot}), and mass-loss rate (\S\ref{sec:massloss}) by comparing with models involving a single star. In \S\ref{sec:heii4686} we show that the observed profile of He\,{\sc ii}\,$\lambda$4686 can not be reproduced with a single star. In \S\ref{sec:luminosity} we will discuss the luminosity constraint that follows from $M_V$.

\begin{figure}
\includegraphics[width=8.5cm]{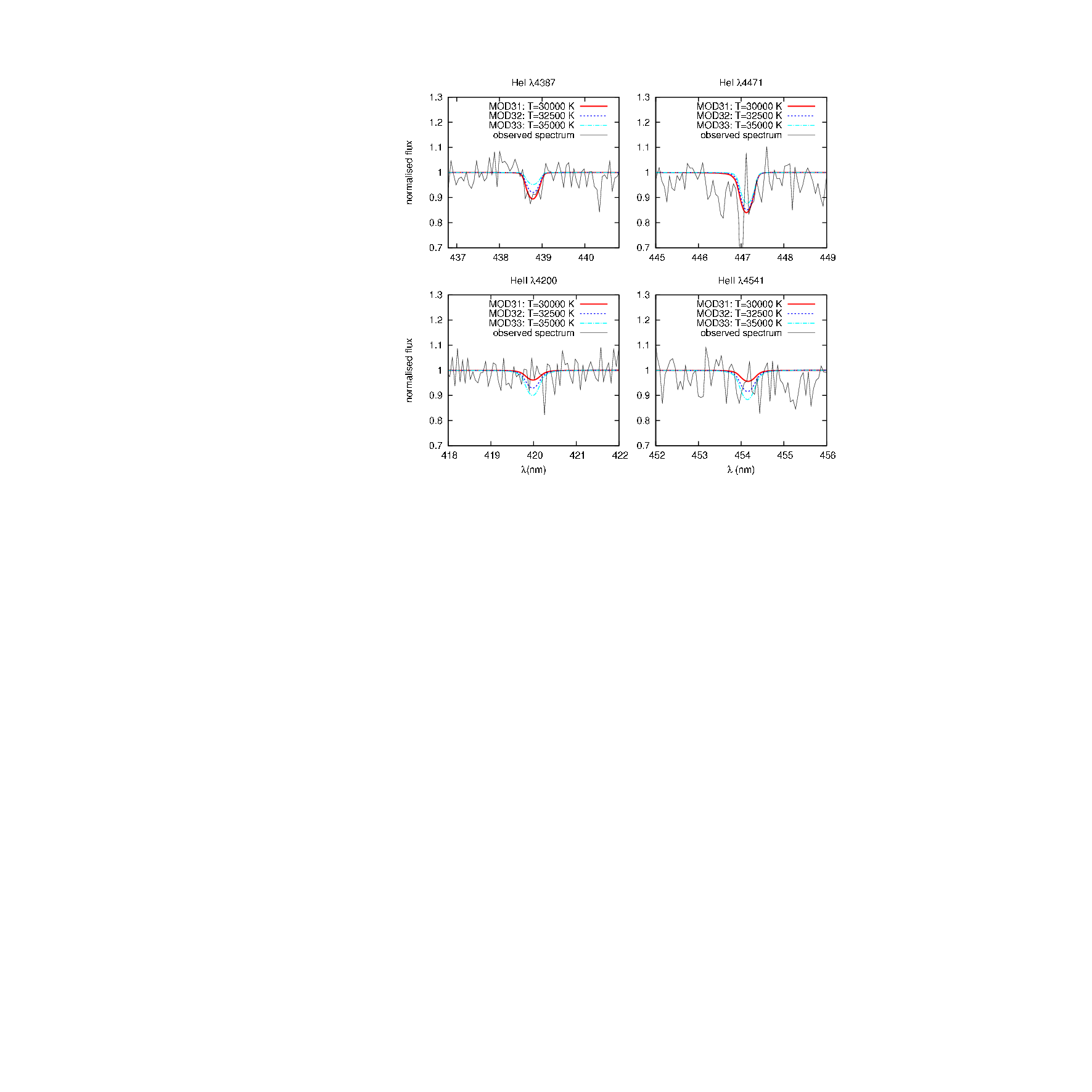}

\caption{Observed line profiles compared with model line profiles computed for various effective surface temperatures (see text).  
\label{fig:plotslineprofiles}}
\end{figure}

\subsubsection{Effective temperature}
\label{sec:modeltemp}
The He\,{\sc i} and He\,{\sc ii} lines can be used to determine the characteristic effective temperature of the source. Fig.~\ref{fig:plotslineprofiles} shows the observed profiles of He\,{\sc i}\,$\lambda$4387, He\,{\sc i}\,$\lambda$4471, He\,{\sc ii}\,$\lambda$4200 and He\,{\sc ii}\,$\lambda$4541 along with profiles from atmosphere models that only differ in effective temperature. A visual comparison of the profiles shows that the models with $T_{\mathrm{eff}}< 35\,000$ K best reproduce the spectrum as otherwise He\,{\sc ii} would have been detected, and He\,{\sc i}\,$\lambda$4387 would not have been as deep.

\subsubsection{Rotational velocity}
\label{sec:vrot}
\begin{figure}
\center
\includegraphics[width=6cm]{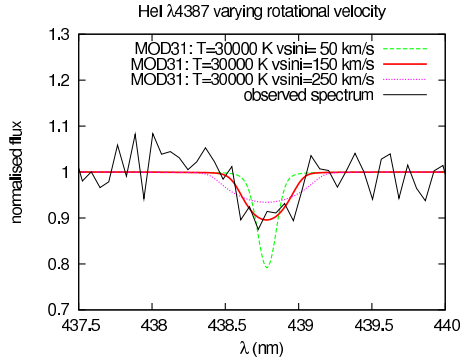}
\caption{The observed line profile of He\,{\sc i}\,$\lambda$4387 compared with the same model profile computed for various projected rotational velocities.\label{fig:vrot}}
\end{figure}

Since the observed profile of He\,{\sc i}\,$\lambda$4387 is not affected by nebular emission, we can use it to constrain the characteristic rotational broadening. Fig.~\ref{fig:vrot} shows the He\,{\sc i}\,$\lambda$4387 profile of MOD31 ($T_{\mathrm{eff}}=30\,000$ K) convolved with rotational profiles to simulate three different rotational velocities: $v_{\mathrm{rot}} \sin({i})=$ 50, 150 and 250~km~s$^{-1}$. The line with $v_{\mathrm{rot}} \sin({i})= 50~\mathrm{km~s}^{-1}$ is clearly not broad enough to fit the observed profile, while the line with $v_{\mathrm{rot}} \sin({i})= 250~\mathrm{km~s}^{-1}$ appears to be too broad. We conclude that the width of the lines is best reproduced by models with $v_{\mathrm{rot}} \sin({i})=150 \pm 50$~km~s$^{-1}$.

\subsubsection{Mass-loss rate}
\label{sec:massloss}
In the nebular spectrum H$\alpha$ is very strongly in emission. This is mostly due to the surrounding H\,{\sc ii} region. In the sky-corrected object spectrum (Fig.~\ref{fig:halpahe}), the H$\alpha$ line wings are clearly visible, and reveal that H$\alpha$ is in emission in the spectrum of C1\_31 as well. \\

The profile of H$\alpha$ is very sensitive to the mass-loss flux $\dot{M}/4\pi R_*^{2}$. In the right panel in Fig.~\ref{fig:halpahe} we plot the observed spectrum together with the line profiles from single-star models with  $T_{\mathrm{eff}}=30\,000$~K and $R_*=23.4\,\mathrm{~R}_{\odot}$, and four different mass-loss rates: $1\times 10^{-5}$, $6\times 10^{-6}$, $3\times 10^{-6}$, and $1\times 10^{-6} \mathrm{~~M}_{\odot}\mathrm{~yr}^{-1}$. Depending on the mass-loss rate, the line is either in emission or in absorption. The computed profile for MOD31 reproduces the observed profile best, and corresponds to a mass-loss flux close to $\sim$$9 \times10^{-10}\mathrm{~ M}_{\odot}$~yr$^{-1}\mathrm{~ R}_{\odot}^{-2}$ (i. e. $\dot{M}=6\times 10^{-6} \mathrm{~~M}_{\odot}\mathrm{~yr}^{-1}$ for $R_*=23.4\mathrm{~R}_{\odot}$).

\subsubsection{He\,{\sc ii}\,$\lambda$4686}
\label{sec:heii4686}
In the spectrum of C1\_31, He\,{\sc ii}\,$\lambda$4686 is in emission, which is a common feature in spectra of O-type supergiants (see Fig.~\ref{fig:compare_obj}). However, in none of the comparison spectra this line is as broad as in our spectrum ($\sim$$3000$ km s$^{-1}$). The equivalent width is only $-3.6\pm0.04$\,\AA, i.e. much weaker than in typical WN star spectra. The feature is of stellar origin as no nebular counterpart is detected. \\

\begin{figure}
\includegraphics[width=8.5cm]{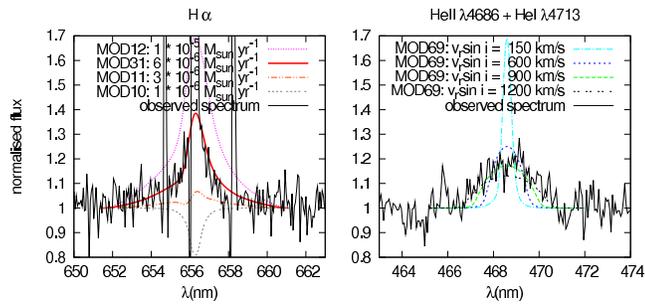}
\caption{Observed line profiles compared with model line profiles for various mass-loss rates (left panel) and projected rotational velocities (right panel).
\label{fig:halpahe}}
\end{figure}

The strength of He\,{\sc ii}\,$\lambda$4686 depends on various parameters in the FASTWIND models. The modeled line can be made stronger by (1) increasing the effective temperature (a larger fraction of the helium will be ionised), (2) increasing the mass-loss rate, (3) increasing the helium abundance, (4) increasing the wind acceleration parameter $\beta$ or (4) decreasing the terminal wind velocity. But these modifications only make the line stronger, not much broader. To simulate the profile of the line in both strength and width using rotational broadening, a rotational velocity of $v_{\mathrm{rot}} \sin({i})=$ 1200~km~s$^{-1}$ is needed: Fig.~\ref{fig:halpahe} shows one of the models (MOD69) for which this line has an equivalent width of $-3.58\,\AA$, convolved with a rotational profile simulating rotational velocities of $v_{\mathrm{rot}}\sin({i})=150$, 600, 900 and 1200\,km\,s$^{-1}$. Only when $v_{\mathrm{rot}}\sin({i})\sim1200$\,km s$^{-1}$, the line is broad enough to reproduce the observed profile. This $v_{\mathrm{rot}}$ is, for acceptable radii and masses, higher than the escape velocity. He\,{\sc ii}\,$\lambda$4686 is formed in the wind, therefore its width can be much higher than the surface rotational velocity. But with such a dense and fast wind, H$\alpha$ would have been much broader as well, and the photospheric He\,{\sc i} and He\,{\sc ii} absorption lines would not be visible. Therefore we conclude that the He\,{\sc ii}\,$\lambda$4686 emission line has a different origin: it is a diluted WN feature (see Section~\ref{sec:consistentpic}).

\subsubsection{Luminosity}
\label{sec:luminosity}
In the calibration of O-stars by \citet{martins2005} the visually brightest supergiant has $M_V=-6.35$. C1\_31 is with $M_V=-8.7$ almost an order of magnitude brighter. It is therefore likely that C1\_31 is a composite object such as a cluster containing several luminous stars. The slit width ($0.8''$ for UVB) at a distance of 2.0~Mpc corresponds to a physical size of 7.8~pc, which is large enough to contain for example the Orion Trapezium cluster, or even more massive open clusters such as Tr~14 \citep{sana2010} or NGC~6231 \citep{sana2008}.  \\

In summary, the normalised line profiles of hydrogen and helium, except He\,{\sc ii}\,$\lambda$4686, can be reproduced by a single-star model with parameters $T_{\mathrm{eff}}\sim 30\,000$ K and $\dot{M} \sim 6 \times 10^{-6}\mathrm{~ M}_{\odot}$~yr$^{-1}$, $R_*\sim23.4\,\mathrm{R}_{\odot}$ and $v_{\mathrm{rot}}\sin({i})=150 \pm 50$~km~s$^{-1}$, i.e. a late O supergiant star (MOD31 in Table~\ref{tab:modparam}). A star with this temperature would need a luminosity of $\log(L_*/L_{\odot})\sim6.3$ to reproduce $M_V$, which is too high for a single O-type supergiant. The considerations regarding the He\,{\sc ii}\,$\lambda$4686 line also point in the direction of the spectrum being a composite of different sources, weighed by their visual brightness. This scenario will be explored in Section~\ref{sec:consistentpic}.

\section{Analysis: The nebular emission line spectrum}
\label{sec:surrh2}

\begin{table}
\caption{Reddening corrected (see text) line flux ratios with respect to H$\beta$ for the strongest unblended emission lines in the nebular spectrum close to our central source. 
\label{tab:lineratio}}
\begin{tabular}{l@{$\,\lambda$}r l@{\,$\pm$\,}l l@{\,$\pm$\,}l l@{\,$\pm$\,}l}

\multicolumn{2}{l}{Line}&\multicolumn{2}{c}{13/08/09}	&	\multicolumn{2}{c}{27/09/09}		&	\multicolumn{2}{c}{Average}	\\
\multicolumn{2}{l}{}&\multicolumn{2}{c}{Ratio}	&	\multicolumn{2}{c}{Ratio}		&	\multicolumn{2}{c}{Ratio}	\\
\hline																
[O\,{\sc ii}]	&	3726.0	&	1.554	&	0.050	&	1.476	&	0.033	&	1.500	&	0.028	\\
		
[O\,{\sc ii}]	&	3728.8	&	2.271	&	0.065	&	2.188	&	0.044	&	2.214	&	0.036	\\

H-9	&	3835.4	&	0.062	&	0.011	&	0.060	&	0.007	&	0.061	&	0.006	\\

[Ne\,{\sc iii}]	&	3868.8	&	0.325	&	0.018	&	0.306	&	0.012	&	0.312	&	0.010	\\

	H-8	&	3889.1	&	0.201	&	0.014	&	0.198	&	0.009	&	0.199	&	0.008	\\

H$\delta$	&	4102.0	&	0.275	&	0.014	&	0.271	&	0.010	&	0.272	&	0.008	\\

H$\gamma$	&	4340.5	&	0.517	&	0.019	&	0.505	&	0.013	&	0.509	&	0.011	\\

[O\,{\sc iii}]	&	4363.2	&	0.039	&	0.004	&	0.037	&	0.003	&	0.037	&	0.002	\\

He\,{\sc i}	&	4471.0	&	0.034	&	0.005	&	0.034	&	0.004	&	0.034	&	0.003	\\

He\,{\sc ii}	&	4686.0	&	0.006	&	0.002	&	0.003	&	0.001	&	0.003	&	0.001	\\

H$\beta$	&	4861.3	&	1.000	&	0.027	&	1.000	&	0.019	&	1.000	&	0.016	\\

[O\,{\sc iii}]	&	4958.9	&	1.171	&	0.029	&	1.144	&	0.020	&	1.153	&	0.016	\\

[O\,{\sc iii}]	&	5006.7	&	3.477	&	0.074	&	3.426	&	0.052	&	3.443	&	0.043	\\

He\,{\sc i}	&	5876.0	&	0.110	&	0.007	&	0.113	&	0.005	&	0.112	&	0.004	\\

[S\,{\sc iii}]	&	6312.0	&	0.015	&	0.002	&	0.017	&	0.001	&	0.016	&	0.001	\\
	
[N\,{\sc ii}]	&	6548.0	&	0.060	&	0.004	&	0.062	&	0.003	&	0.061	&	0.002	\\

H$\alpha$	&	6562.8	&	3.040	&	0.063	&	3.052	&	0.045	&	3.048	&	0.037	\\
	
[N\,{\sc ii}]	&	6583.4	&	0.179	&	0.007	&	0.193	&	0.005	&	0.188	&	0.004	\\
	
He\,{\sc i}	&	6678.0	&	0.029	&	0.003	&	0.028	&	0.002	&	0.028	&	0.002	\\
	
[S\,{\sc ii}]	&	6716.5	&	0.269	&	0.009	&	0.292	&	0.007	&	0.283	&	0.005	\\
	
[S\,{\sc ii}]	&	6730.8	&	0.189	&	0.007	&	0.206	&	0.005	&	0.200	&	0.004	\\
	
[Ar\,{\sc v}]&	7005.9	&	0.009	&	0.001	&	0.006	&	0.001	&	0.007	&	0.001	\\

[Ar\,{\sc iii}]&	7135.8	&	0.087	&	0.004	&	0.086	&	0.003	&	0.087	&	0.002	\\

[Ar\,{\sc iii}]&	7751.1	&	0.022	&	0.001	&	0.022	&	0.001	&	0.022	&	0.001 	\\	
	
Pa-10	&	9015.0	&	0.017	&	0.001	&	0.018	&	0.001	&	0.018	&	0.001	\\
	
[S\,{\sc iii}]	&	9068.9	&	0.200	&	0.005	&	0.203	&	0.004	&	0.202	&	0.003	\\
	
Pa-9	&	9229.0	&	0.025	&	0.002	&	0.024	&	0.001	&	0.024	&	0.001	\\
	
[S\,{\sc iii}]	&	9531.0	&	0.475	&	0.011	&	0.437	&	0.008	&	0.449	&	0.006	\\
	
Pa-7	&	10049.4	&	0.045	&	0.004	&	0.046	&	0.003	&	0.046	&	0.002	\\
\hline
\end{tabular}
\end{table}

The nebular emission spectra show hydrogen recombination lines of the Balmer and Paschen series, He\,{\sc i} lines, and forbidden lines of O\,{\sc ii}, O\,{\sc iii}, S\,{\sc ii}, S\,{\sc iii}, N\,{\sc ii}, Ne\,{\sc iii}, Ar\,{\sc iii}, Ar\,{\sc iv} and Ar\,{\sc v}. To these spectra, no sky-correction could be applied, because nebular emission covers the full slit. Therefore, every nebular line we measure might have a contribution from the sky continuum. We minimize this error by subtracting the local continuum next to the line in wavelength. The ratios with respect to H$\beta$ of the nebular lines at the position of our source are listed in Table~\ref{tab:lineratio}. The extinction corrected integrated specific intensity of H$\beta$ is $5.8\times 10^{-15}\,\mathrm{erg~s}^{-1} \mathrm{~cm}^{-2}\mathrm{~arcsec}^{-2}$. \\

We use the diagnostics for electron temperature and oxygen abundance from \citet{pagel1992}: 

\begin{equation}
 T  = \frac{1.432}{ \log R - 0.85 +0.03 \log T + \log \left(1+0.0433xT^{0.06}\right) } \,\,,\label{eq:T}
\end{equation}
where $T \equiv T \left( \textrm{O\,{\sc iii}} \right)$, the electron temperature in the region where oxygen is doubly ionized, in units of 
$10^4$\,K. $R$ and $x$ are given by 

\begin{eqnarray}
R		&=		&\frac{I_{\lambda4959}+I_{\lambda 5007}}{I_{\lambda 4363}} \, , \label{eq:R}\\
x		&=		&10^{-4} n_{\mathrm{e}} T_2^{-1/2} \, , \label{eq:x}
\end{eqnarray}
where $I$ is the integrated specific intensity intensity of the indicated emission line and $n_{\mathrm{e}}$ is the electron density in cm$^{-3}$. $T_2$ is the electron temperature in units of 
$10^4$\,K in the singly ionized region, and follows from model calculations by \citet{stasinska1990}:
\begin{eqnarray}
T_2^{-1}	&\equiv	& \left[ T \left( \textrm{O\,{\sc ii},\,N\,{\sc ii},\,S\,{\sc ii}} \right) \right]^{-1} = 0.5\left( T^{-1}+0.8\right) \, . \label{eq:T2}
\end{eqnarray}

\noindent The mean ionic abundance ratios for O{\,\sc ii} and O{\,\sc iii} along the line of sight are calculated as follows:
\begin{eqnarray}
12+\log \left( \textrm{O{\,\sc ii}\,/\,H{\,\sc ii}}	\right)&=&	 \log \frac{I_{\lambda 3726}+I_{\lambda 3729}}{I_{\mathrm{H}\beta}} +5.890 +\frac{1.676}{T_2}\nonumber \\
							& & -0.40 \log T_2 + \log \left( 1+1.35 x \right);  \label{eq:ox1}\\
12+\log \left( \textrm{O{\,\sc iii}\,/\,H{\,\sc ii}}	\right)&=&	 \log \frac{I_{\lambda4959}+I_{\lambda 5007}}{I_{\mathrm{H}\beta}} +6.174 +\frac{1.251}{T}\nonumber \\
							& & -0.55 \log T ; \label{eq:ox2}						
\end{eqnarray}

\noindent The total oxygen abundance is obtained by adding Eqs.~\ref{eq:ox1} and \ref{eq:ox2}, i.e. by assuming these two ionisation stages to be dominant in the H\,{\sc ii} region. \\

\subsection{Nebular emission properties along the slit}
\label{sec:nebprop}
\par
\begin{figure*}
\includegraphics[width=8.7cm]{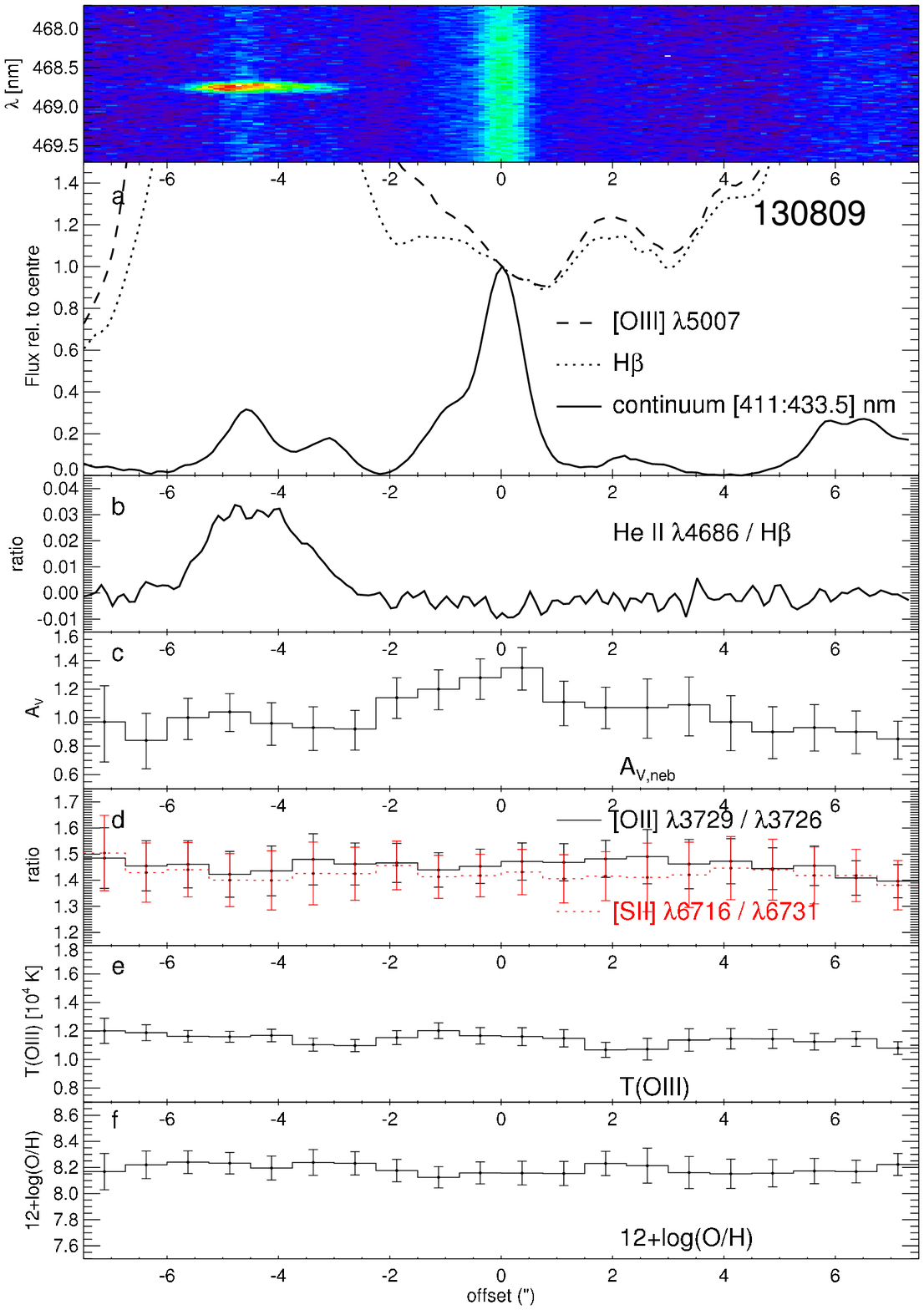}
\includegraphics[width=8.7cm]{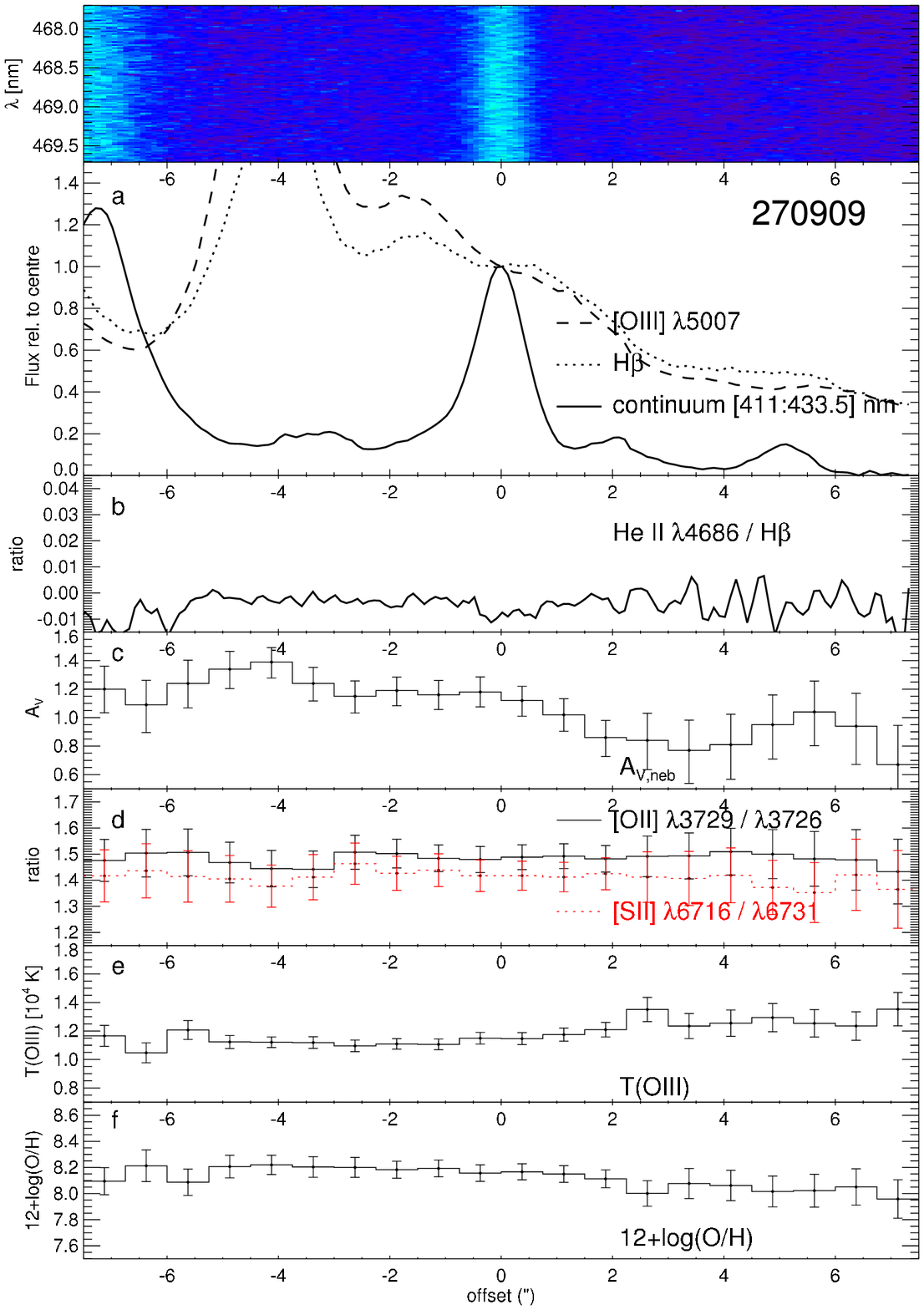}
\caption{Spatial profiles of emission lines and properties measured from the emission line spectrum in 20 sub-apertures along the slit On the x-axis is distance in arcsec from C1\_31. The left and right frames, labeled by their observing dates, correspond to observations with different slit orientations, see Fig.~\ref{fig:regionslit} and Table~\ref{tab:observations}. The image at the top is a part of the 2D spectrum around 468.6\,nm. Panel (a) shows the average level of the continuum between $411.0-433.5$\,nm, without taking into account the difference in $A_{V,\mathrm{neb}}$ along the slit. It also shows the intensity of the two strong nebular lines H$\beta$ and [O\,{\sc iii}] $\lambda5007$, with respect to the value at the location of the central source. Panel (b) shows the flux of the He\,{\sc ii}\,$\lambda$4686 nebular emission line with respect to H$\beta$. Panel (c) shows the value of $A_{V,\mathrm{neb}}$, derived from a number of Balmer and Paschen emission lines, assuming Case B recombination. Panel (d) shows [O\,{\sc ii}] and [S\,{\sc ii}] ratios that are sensitive to electron density. Panel (e) shows $T\left( \textrm{O\,{\sc iii}} \right)$ calculated from the ratio $R$ of the [O\,{\sc iii}] lines (Eq.~\ref{eq:T}). Panel (f) shows the total oxygen abundance $12+\log \left( \mathrm{O} / \mathrm{H} \right)$. }
\label{fig:spatial130809}
\end{figure*}

 Fig.~\ref{fig:spatial130809} shows the nebular properties along the slit for the spectra obtained on 13/08/09 and 27/09/09. The orientation of the slit differs between these observations: positive offset is $\sim$North-East for 13/08/09 and $\sim$North-West for 27/09/09  (see Fig.~\ref{fig:regionslit}). We do not show the results for the 30/09/09 observation, which has a similar position angle and gives a similar result to 27/09/09, though with larger error bars. As mentioned in Section~\ref{sec:rednebspec}, the errors are obtained by propagating the photon noise on the intensity of the used lines.  We choose $n_{\mathrm{e}}= 20\,\mathrm{cm}^{-3}$ arbitrarily, but in agreement with the low density limit of the density-sensitive [O\,{\sc ii}] and [S\,{\sc ii}] ratios we measure (see panel (d) in Fig.~\ref{fig:spatial130809}, and the analysis below). The error on the density is not propagated into the errors on the other parameters, because they all depend very weakly on the density (see Eqs.~\ref{eq:T} and \ref{eq:x}). In the following sections we discuss the panels in Fig.~\ref{fig:spatial130809}. 

\subsubsection*{Stellar continuum and nebular emission}
Panel (a) shows the shape of the continuum between $411.0-433.5$\,nm; here we see the trace of our central source. We do not take into account the difference in $A_V$ along the slit (see panel c).
The nebular emission in H$\beta$ and [O\,{\sc iii}] $\lambda$5007, two of the strongest emission lines in the spectra, is also shown. [O\,{\sc iii}] $\lambda$5007 is a forbidden transition and thus only emitted by low density nebulae.
The 27/09/09 observation suggests that the nebular emission is built up from various discrete peaks, of which one is centered on C1\_31.  This part of the nebula, with a radius of $\sim$$20-30$\,pc, is likely ionized by C1\_31. Therefore, the nebular properties at the central location in the slit are related to the properties of C1\_31 (See Section~\ref{sec:modelhii}).

\subsubsection*{He\,{\sc ii}\,$\lambda$4686\,/\,H$\beta$}
Panel (b) shows the ratio of the He\,{\sc ii}\,$\lambda$4686 and H$\beta$ nebular emission line. Around C1\_31, this nebular line is not present, but it is very pronounced in the 13/08/09 spectrum around offset $-4.5''$. This is also clearly visible in the 2D spectrum around this line (top left image of Fig.~\ref{fig:spatial130809}). This feature will be discussed in more detail in Section~\ref{sec:C235}. This nebular feature is much narrower than the wind He\,{\sc ii}\,$\lambda$4686 line we discussed in Section~\ref{sec:heii4686}. The nebular He\,{\sc ii}\,$\lambda$4686 line has a Gaussian FWHM of $\sim$1\,\AA\, like the other nebular lines, slightly larger than the resolution element at this wavelength ($\sim$0.75\,\AA).

\subsubsection*{Extinction}
Panel (c) shows the values of $A_{V,\mathrm{neb}}$, which is the extinction derived from the ISM hydrogen line ratios. Adopting $R_V=3.24$ \citep{gieren2008}, $A_{V,\mathrm{neb}}$ is found for each aperture by minimizing the following expression:
\begin{equation}
\chi_{\mathrm{red}}^2=\frac{\chi^2}{\nu}=\frac{1}{\nu} \sum_{lines} \frac{(q-q_0)^2}{\sigma_{q}^2}
\end{equation}
with $q_0$ the theoretical ratio for Case B recombination in the low density limit at $T=10\,000$\,K \citep[see e.g.][]{osterbrock2006}, $q$ the intensity of a line with respect to H$\beta$ after applying $A_{V,\mathrm{neb}}$ (following \citealt{cardelli1989}), $\sigma_{q}^2$ the variance on $q$, and $\nu$ the number of degrees of freedom. We use H-9, H$\gamma$, H$\delta$, H$\alpha$, Pa-10, Pa-9 and Pa-7. 
The confidence interval on $A_{V,\mathrm{neb}}$ is given by the value for which $\chi_{\mathrm{red}}^2$ rises by 1. $A_{V,\mathrm{neb}}$ is used in the derivation of the properties per aperture shown in panels (d) to (f) of Fig.~\ref{fig:spatial130809}; the error on $A_{V,\mathrm{neb}}$ is not propagated as its influence on the other parameters is small.\\
$A_{V,\mathrm{neb}}$ at the location of C1\_31 is derived to be $1.30\pm0.15$ and $1.15\pm0.10$ in the 13/08/09 and 27/09/09 spectra respectively. This is lower than $A_{V,\mathrm{star}}=2.3\pm0.1$ (see Section~\ref{sec:photospec} and Fig.~\ref{fig:fullspec_dered}). 
Furthermore, we note that both $A_{V,\mathrm{neb}}$ and $A_{V,\mathrm{star}}$ are larger than $A_V=0.45$ from \citet{gieren2008}. However, this latter value is an average for NGC\,55 as a whole. This range of values reflects local variations and are to be expected, especially in an almost edge-on galaxy.\\
\subsubsection*{Electron density}
Panel (d) shows [O\,{\sc ii}] and [S\,{\sc ii}] ratios that are sensitive to electron density \citep[see e.g.][]{osterbrock2006}. For a value $>1.4$, both ratios are in the low density ($<10^2$\,cm$^{-3}$) limit, so this measurement only provides an upper limit. Because these line pairs are very close in wavelength, their ratios are not affected by $A_{V,\mathrm{neb}}$.

\subsubsection*{Electron temperature}
Panel (e) gives the electron temperature calculated from the ratio $R$ of the [O\,{\sc iii}] lines (Equation~\ref{eq:T}). In the 27/09/09 observation (Fig.~\ref{fig:spatial130809}, right) in the apertures with $\mathrm{offset}>2''$, [O\,{\sc iii}] $\lambda 4363$ was hardly detected resulting in an underestimation of the error in $T \left( \textrm{O\,{\sc iii}} \right)$. Taking this into account, and weighing the better quality spectra more strongly, we conclude that around the location of C1\_31 $T \left( \textrm{O\,{\sc iii}} \right) = 11500\,\pm\,600$\,K, and that there are no significant gradients in the two spatial directions indicated in Fig.~\ref{fig:regionslit}. 

\subsubsection*{Oxygen abundance}
Panel (f) shows the total oxygen abundance $[\mathrm{O} / \mathrm{H}]=12+\log \left( \mathrm{O} / \mathrm{H} \right)$. The slight drop in [O/H] that we see at $\mathrm{offset}>2''$ in the 27/09/09 observation is a propagated effect from the uncertain determination of $T \left( \textrm{O\,{\sc iii}} \right)$. Excluding this region, we find an average of $[\mathrm{O} / \mathrm{H}] = 8.18 \pm 0.03$, which corresponds to $Z=0.31 \pm 0.04\,Z_{\odot}$ adopting $[\mathrm{O}/ \mathrm{H}]_{\odot} = 8.69 \pm 0.05$ \citep{asplund2009}. Though lower oxygen abundances are reported for NGC\,55 (\citealp[$8.08 \pm 0.10$,][]{tuelmann2003}), on average slightly higher values are measured (\citealp[$8.23-8.39$,][]{webster1983}; \citealp[8.53,][]{stasinska1986}; \citealp[$8.35\pm0.07$,][]{zaritsky1994}).

\subsection{C2\_35}
\label{sec:C235}
C2\_35 (RA 0:14:59.68, DEC -39:12:42.84) is located $4.5''$ West South-West of C1\_31 (see Fig.~\ref{fig:regionslit}). Its strongly ionised surrounding nebula is visible in the 13/08/09 spectrum (Fig.~\ref{fig:spatial130809}, left, panel a). At offset $-4.5''$ we see a weak continuum, as the point-spread function of C2\_35 is mostly outside the slit. We detect strong  (forbidden) nebular line emission (panel a), but the most striking feature is the high He\,{\sc ii}\,$\lambda$4686\,/\,H$\beta=0.03$ ratio (left, panel~b and top image).
The FORS2 spectrum of C2\_35, classified as an early OI by \citet{castro2008}, is similar to C1\_31. The broad He\,{\sc ii}\,$\lambda$4686 wind feature is indicative of the presence of a hot WR star. On top of the broad wind feature, there is a narrow nebular emission line. 
He\,{\sc ii} emission lines from nebulae are only rarely seen and often associated with strong X-ray sources \citep[see e.g][]{pakull1986,kaaret2004}, but see \citet{shirazi2012}. No obvious X-ray source is detected in archival XMM-Newton and Chandra observations at the location of  C2\_35 (priv. comm. R. Wij\-nands). The exposure times of these images, however, would not be sufficient to detect the X-ray emission of, for example, a stellar mass black hole at this distance. \\

\section{Model H\,{\sc ii} region}
\label{sec:modelhii}

\begin{figure}
\includegraphics[width=8.5cm]{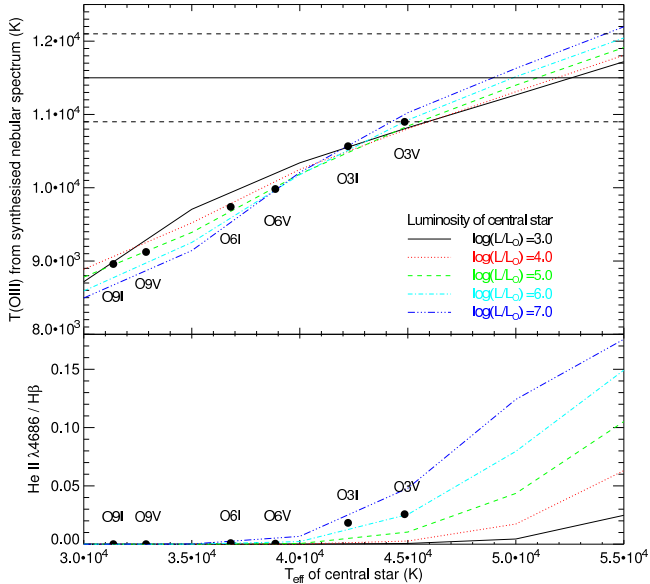}
\caption{Upper panel: the predicted $T$(O\,{\sc iii}) from the synthesized nebular spectrum as a function of $T_{\mathrm{eff}}$ and $L_*$ of a central ionizing O-star \citep{lanz2003} in a CLOUDY model with an inner radius of 0.1~pc. The solid and dashed horizontal lines are the observed $T$(O\,{\sc iii}) and error estimate for the direct environment of C1\_31. Lower panel: the predicted nebular line ratio He\,{\sc ii}\,$\lambda\,4686$/H$\beta$ as a function of $T_{\mathrm{eff}}$ and $L_*$ of a central ionising star. The black dots correspond to realistic dwarf and supergiant O-type stars in the $T_{\mathrm{eff}}$ and $L_*$ grid \citep{martins2005}.}
\label{fig:CLOUDY_TO3}
\end{figure}

If the stellar spectrum is a composition of different sources, various solutions are possible. However, the properties $T$(O\,{\sc iii}), $n_{\mathrm{e}}$, and [O/H] of the surrounding nebula are constrained (see Section~\ref{sec:surrh2}), and the nebular emission profiles along the slit give some idea of the size of the ionised region. The ionising source, of which the constituents are constrained by the stellar spectrum, should be able to produce a region with properties we derive from the nebular spectrum. In this section we will use the spectral synthesis code CLOUDY (version 08.00, \citealt{ferland1998}) to investigate which properties of the nebula and of the central ionising source have a strong influence on the observables that we measured. This will allow us to constrain $T_\mathrm{eff}$ and $L_*$ of the ionising source.\\

CLOUDY is designed to simulate gaseous interstellar media. From a given set of conditions such as luminosity and spectral shape of the ionizing source, the program computes the thermal, ionisation and chemical structure of a region as well as the emitted spectrum. In order to compare our model results to what has been observed, we will mainly use the simulated emitted spectrum.
Since we only know the total abundance of oxygen, we use the abundance pattern corresponding to the Orion Nebula provided by CLOUDY \citep{baldwin1991,rubin1991,osterbrock1992,savage1996}, and scale the metallicity such that the oxygen abundance matches our observed value. The total hydrogen density $n_\mathrm{H}$ is set to $20$~cm$^{-3}$, consistent with the low electron density limit derived from [O\,{\sc ii}] and [S\,{\sc ii}], assuming that all hydrogen is ionised. We do not include dust grains\footnote{We exclude dust grains to keep the model simple. The extinction we measure both in the stellar spectrum and the hydrogen emission lines could as well be due to dust that is outside the surrounding ionised region.}. We use a spherical geometry: the inner radius of the cloud is 0.1 pc and the outer radius is set where the temperature drops below 4\,000\,K. \\

We examined the ratio $R$ (Eq.~\ref{eq:R}) of the modeled output spectra, and computed $T$(O\,{\sc iii}), as a function of $T_{\mathrm{eff}}$ and $L_*$ of a synthetic O-star spectrum \citep{lanz2003} as central ionising source; see the upper panel of Fig.~\ref{fig:CLOUDY_TO3}. A grid of $T_{\mathrm{eff}}=30\,000-55\,000$\,K and $\log(L/L_{\odot})=3-6$ has been examined; the black dots indicate realistic stars in this grid according to the calibration by \citet{martins2005}.
We see that a higher $T_{\mathrm{eff}}$ of the central star leads to a higher $T$(O\,{\sc iii}), while the effect of $L_*$ on the `measured' $T$(O\,{\sc iii}) is smaller. $L_*$ does however influence the size of the cloud: a higher $L_*$ leads to a larger cloud. A cloud with metallicity $Z=0.3\,\mathrm{ Z}_{\odot}$ needs a $T_{\mathrm{eff}}\sim44\,000-55\,000$ K star, depending on the luminosity, to produce the measured  $T$(O\,{\sc iii}) of $11\,500\pm600$\,K (horizontal line). Furthermore, we find that the metallicity of the cloud has an even stronger influence on $T$(O\,{\sc iii}); lower metallicities result in hotter clouds because they are less efficiently cooled. Changing $n_\mathrm{H}$ only influences the spatial scale of the cloud: a density ten times as low results in a region four times as large. \\

We also analysed the nebular line ratio He\,{\sc ii}\,$\lambda\,4686$/H$\beta$ as a function of $T_{\mathrm{eff}}$ and $L_*$ of a central ionising source; see the lower panel of Fig.~\ref{fig:CLOUDY_TO3}. Below $T_{\mathrm{eff}}\simeq40\,000$\,K, no significant He\,{\sc ii}\,$\lambda\,4686$ line is predicted for any of the luminosities in our grid. For  $T_{\mathrm{eff}}>40\,000$\,K, the line can be produced, and is stronger with respect to H$\beta$ for more luminous sources. The ratio He\,{\sc ii}\,$\lambda\,4686$/H$\beta$ decreases strongly by increasing the inner radius of the model cloud. The nebula directly around C1\_31 does not show He\,{\sc ii}\,$\lambda\,4686$, therefore the inner radius is likely to be 1~pc or more. $T$(O\,{\sc iii}) is not affected significantly by changing the inner radius.

\section{Discussion: a consistent picture}
\label{sec:consistentpic}

\begin{figure}
\includegraphics[width=8.3cm]{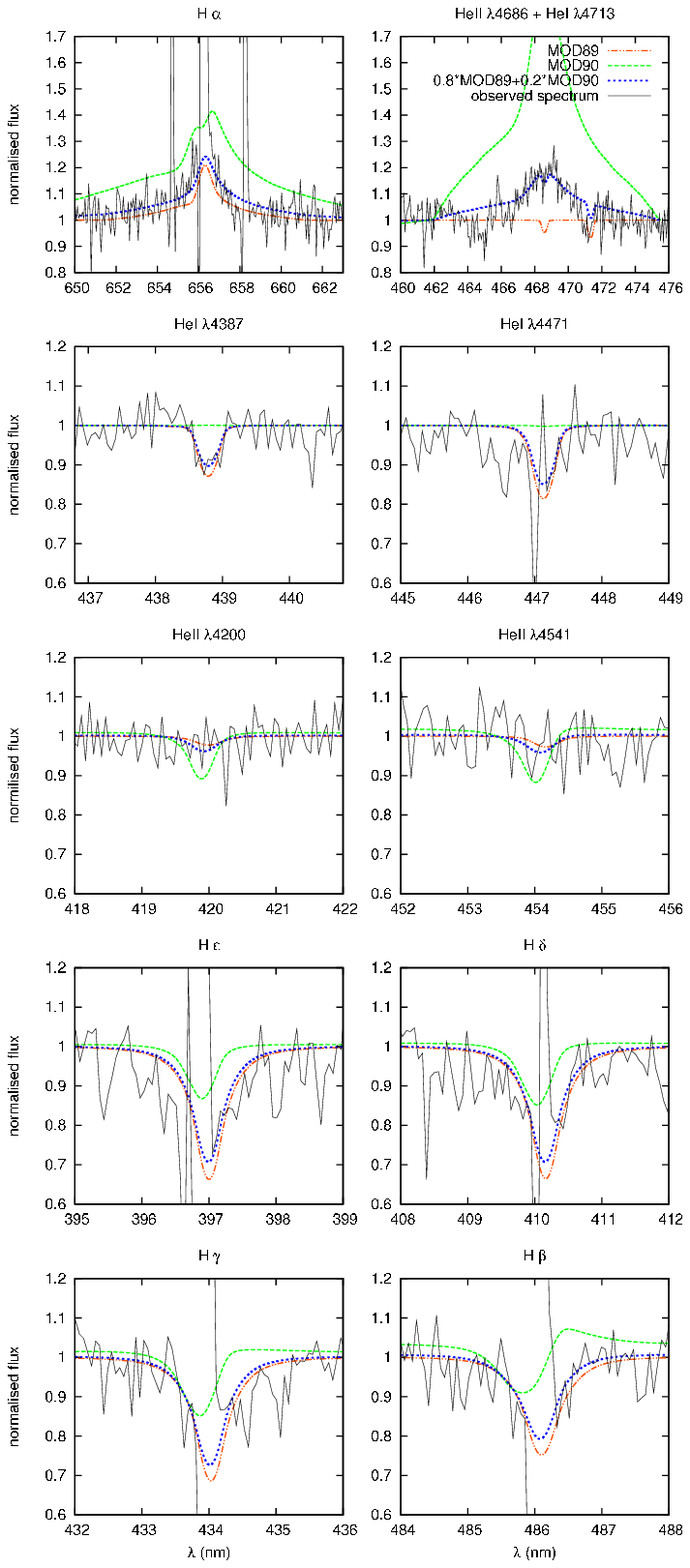}
\caption{The observed normalised spectrum (solid lines) together with the combined line profiles (dotted lines) from the models described in Table~\ref{tab:modparam}, representing the late O giant component MOD89 (dash-dot-dot lines) and the WN-like component MOD90 (dashed lines), in the ratio $4:1$.}
\label{fig:combinedmodelspec}
\end{figure}

\begin{figure}
\includegraphics[width=8.3cm]{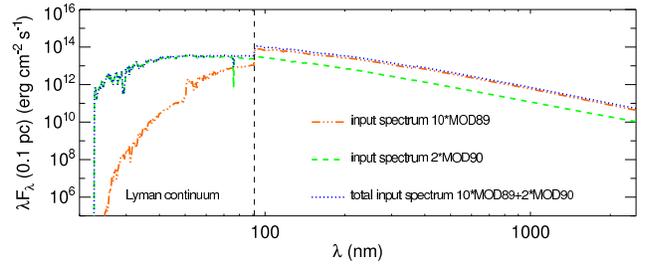}
\caption{The spectral energy distribution at the inner radius of the model cloud ($r_{\mathrm{in}}=0.1$\,pc), split in the different components. It is clear that in the visual, the 30\,000\,K component (MOD89, dot-dashed lines) dominates the light, while the hydrogen ionising photons (left of the vertical dashed line, with $hv>13.6$\,eV) are provided mainly by the hot (50\,000\,K) WN component (MOD90, dashed lines). The dotted profile shows the total input spectrum.
\label{fig:fullspec_cloudy}}
\end{figure}

In the previous sections, we have put constraints on the stellar parameters of C1\_31 and its surrounding region. We suggest that C1\_31 is not a single object, but rather a stellar cluster. We summarise the main arguments below.
\begin{enumerate} 
\item[(a)] the line profiles can not be reproduced by one single stellar atmosphere model, especially not He\,{\sc ii}\,$\lambda$4686 (Section~\ref{sec:heii4686});
\item[(b)] the visual absolute magnitude $M_V=-8.7$ of this source is very high for a single object (Section~\ref{sec:photospec});
\item[(c)]a very hot central object ($\sim$50\,000\,K) is necessary to produce a $T$(O\,{\sc iii}) of $\sim$11\,500\,K in the surrounding nebula (Section~\ref{sec:modelhii}), but the ''average'' spectral type suggests that the majority of the luminosity in the visual is produced by $T_{\mathrm{eff}}\lesssim35\,000$ K stars (Section~\ref{sec:modeltemp}).
\end{enumerate}

The spectrum of a composite object results in a superposition of all components, weighed by their relative brightness at the wavelength considered. All lines that we analysed are close in wavelength, so we adopt a general ratio in brightness which is the ratio in $V$. Given a universal initial mass function (IMF), a cluster will consist of many cool low-mass stars, and only a few very luminous and hot ones. In this analysis we will focus on the most massive and luminous members, because these dominate the cluster spectrum, as well as the flux of ionising photons. To this end we present a simple combination of FASTWIND models that (a) reproduces all observed line profiles, (b) has an absolute magnitude $M_V=-8.7$, and (c) when put into a CLOUDY model of an H\,{\sc ii} region, produces an electron temperature $T$(O\,{\sc iii})\,$\sim11\,500$\,K, for an adopted metallicity $Z=0.3\,\mathrm{Z}_{\odot}$ and hydrogen density $n_\mathrm{H}=20$\,cm$^{-3}$.\\

\par Model MOD90 (see Table~\ref{tab:modparam}) has a high temperature ($T_{\mathrm{eff}}=50\,000$~K), a high mass-loss rate ($3\times10^{-5}\,\mathrm{M}_{\odot}\mathrm{~yr}^{-1}$) and an enhanced helium abundance  $N_{\mathrm{He}}/N_{\mathrm{H}}=0.8$. It mimics a Wolf-Rayet WN star \citep[see e.g.][]{crowther2008}. These properties result in a strong and broad He\,{\sc ii}\,$\lambda$4686 emission feature (Fig.~\ref{fig:combinedmodelspec}, upper right). To reproduce the observed shape of He\,{\sc ii}\,$\lambda$4686, we combine this profile with a model with a weak He\,{\sc ii}\,$\lambda$4686 absorption profile: model MOD89, with $T_{\mathrm{eff}}=30\,000$~K. This model resembles a late-O/early-B giant or bright giant. We create a combined profile of 20\% MOD90 and 80\% MOD89. With this flux ratio, the He\,{\sc i} and He\,{\sc ii} absorption lines resemble a 30\,000~K star, because in this respect the MOD89 model is dominant. The H$\alpha$ wings are well reproduced (Fig.~\ref{fig:combinedmodelspec}, upper right). The other Balmer lines in the WN component MOD90 are also affected by the strong wind, which results in a shallower line or even a P-Cygni profile in the case of H$\beta$ and H$\gamma$. However, the strong absorption profile of the late O giant component MOD89 dominates, and the combined profiles match the observed ones.\\

If the cluster consists of ten stars like MOD89 and two like MOD90, the ensemble would have a visual magnitude of $M_V=-8.63$, in agreement with the observed value $M_V=-8.7\pm0.4$. The flux ratio in the visual would be $\sim4:1$, due to the different bolometric corrections (see Table~\ref{tab:mv}). The visual flux is dominated by the late O-type component, while further to the UV, the hot Wolf-Rayet component would dominate. The latter is required to reproduce the observed electron temperature in de cloud. \\

\begin{table}
\caption{Properties of the final composition of the cluster. Our model cluster has two different components: a late O giant component (MOD89) and a Wolf Rayet WN-type component (MOD90), see also Table~\ref{tab:modparam}. Model ID, effective temperature, bolometric luminosity, bolometric correction, absolute visual magnitude, number of stars of this type in the cluster, total absolute magnitude of this component, and fraction of the total visual flux provided by this component. \label{tab:mv}}
\begin{tabular}{lccc@{ } ccc@{ }c}
\hline

ID	& $T_{\mathrm{eff}}$	&$L_*$					& BC		& $M_V$	& 	\#	&	$M_{V}$ & Fraction \\	
	&(K)				&$(\mathrm{L}_{\odot})$	&		&		&		&	total		 & of	$F_{\mathrm{vis}}$\\
\hline
MOD89	& 	30\,000	&$10^{5.24}$			& $-2.42$	& $-5.93$	&	10	&	$-8.43$	&	0.8 \\
MOD90	& 	50\,000	&$10^{5.83}$			& $-3.90$	& $-5.95$	&	2	&	$-6.70$	&	0.2 \\

\hline
\end{tabular}
\end{table}

We have used the combined spectrum of 10 times MOD89 and 2 times MOD90 as ionising source in a CLOUDY model. The SED at the inner radius of the cloud is shown in Fig.~\ref{fig:fullspec_cloudy}, where we see that the ionising flux is indeed dominated by the WN component MOD90. We use the same configuration as we did in Section~\ref{sec:modelhii}, with $Z=0.3\,\mathrm{Z}_{\odot}$. We choose $n_{\mathrm{H}}=20$~cm$^{-3}$, such that CLOUDY produces an ionised region with a radius of $\sim$20\,pc, see Section~\ref{sec:nebprop}. From the synthesized nebular spectrum we infer $T$(O\,{\sc iii}) $\sim10\,800$\,K, slightly lower than the measured $T$(O\,{\sc iii}) $\sim11\,500\pm600$\,K, but in reasonable agreement. According to the model in Section~\ref{sec:modelhii} and its results in Figure~\ref{fig:CLOUDY_TO3}, a 50\,000~K star would be able to heat the cloud to $\sim$11\,500\,K. However, adding more late O stars decreases $T$(O\,{\sc iii}).
The predicted nebular line ratio He\,{\sc ii}\,$\lambda\,4686$/H$\beta$ for this cluster composition is 0.06, which challenges the non-detection of nebular line He\,{\sc ii}\,$\lambda\,4686$ around C1\_31. This disagreement can be reconciled by increasing the inner radius of the model cloud to 1~pc or more.\\

In principle, information as to the stellar content may also be derived from considering the mass-loss rates of the contributing stars. For our late O\,II/III source (MOD89) the adopted mass-loss rate of $2\times10^{-6}\mathrm{~ M}_{\odot}$~yr$^{-1}$ is rather large compared to theoretical expectations for such a star at a metallicity of $0.3\,\mathrm{Z}_{\odot}$, being an order of magnitude higher than predicted by \citet{vink2001}, after correcting the empirical mass-loss rate for wind inhomogeneities \citep{mokiem2007}. Using the Vink et al. prescription for the WN star (MOD90) yields a much smaller discrepancy of a factor $\sim$2. 
The discrepancy in the O star mass-loss rate can be partly reconciled by taking a smaller number of brighter stars, for instance supergiants, in the following way: A brighter star has a larger radius. In order to preserve the H$\alpha$ profile shape, the quantity $Q \propto \dot{M} / (R_*^{3/2} v_{\infty})$ needs to remain invariant \citep{dekoter1998}. This implies that for fixed temperature $\dot{M} \propto L_*^{3/4}$. However, the expected mass-loss rate scales as $\dot{M} \propto L_*^{2.2}$, therefore a smaller number of brighter O stars may still match the strong H$\alpha$ emission line wings and better reconcile observed and theoretical mass-loss rates. 
We did not pursue this strategy in view of the uncertainties that are involved, for instance those relating to the stellar mass (which enters the problem as mass-loss is expected to scale with mass as $\dot{M} \propto M^{-1.3}$). Moreover, we remark that higher than expected mass-loss rates have been reported for O stars in low-metallicity galaxies \citep{tramper2011}.

\section{Summary and conclusions}
\label{sec:conclusion}

We have analysed the VLT/X-shooter spectrum of C1\_31, one of the most luminous sources in NGC\,55, and its surroundings. We conclude that NGC\,55\,C1\_31 is a cluster consisting of several massive stars, including at least one WN star, of which we observe the integrated spectrum. \\

The H, He\,{\sc i} and He\,{\sc ii} lines in the stellar spectrum have been compared to synthesized spectra from a grid of FASTWIND non-LTE stellar atmosphere models. All normalised lines except He\,{\sc ii}\,$\lambda\,4686$ can be reproduced by a single-star model with $T_{\mathrm{eff}} \lesssim 35\,000$\,K, $\dot{M} \sim2\times10^{-6}\mathrm{~ M}_{\odot}$~yr$^{-1}$, and $v_{\mathrm{rot}}\sin(i)=150 \pm 50$~km\,s$^{-1}$. He\,{\sc ii}\,$\lambda\,4686$ has an equivalent width of $-3.6\pm0.4$\,\AA, but is $\sim$3000\,km\,s$^{-1}$ wide. No single star model is able to produce matching profiles for all lines simultaneously.\\

Analysis of the nebular emission spectrum along the slit yields an electron density $n_{\mathrm{e}}\leq10^2$\,cm$^{-3}$, electron temperature $T \left( \textrm{O\,{\sc iii}} \right) = 11\,500\,\pm\,600$\,K, and oxygen abundance $[\mathrm{O}/\mathrm{H}]=8.18 \pm 0.03$, which corresponds to a metallicity $Z=0.31\pm0.04\,\mathrm{Z}_{\odot}$. A grid of CLOUDY models suggests that a hot ($\sim$50\,000\,K) ionising source is necessary to reproduce the observed $T \left( \textrm{O\,{\sc iii}} \right)$ in a H\,{\sc ii} region with comparable density and metallicity.\\

We have also presented an illustrative cluster composition that reproduces all observed spectral features, the visual brightness of the target, and which is able to maintain an H\,{\sc ii} region with properties similar to those derived from the nebular spectrum. In our model, the cluster contains several blue (super)giants and one or more WN stars. While the proposed composition might not be unique, the presence of at least one very hot, helium rich star with a high mass-loss is a robust conclusion. High angular resolution imaging reaching a resolution of 0.05\arcsec\ (corresponding to a physical distance of about 0.5~pc) would provide an improvement of a factor 10 to 20 compared to our seeing-limited observations and would help to constrain the composition of the cluster. This makes NGC\,55\,C1\_31 a prime target for ELT-class telescopes combining high angular resolution and integral field or multi-object spectroscopy.

\section*{Acknowledgments}
We acknowledge the X-shooter Science Verification team. We thank Andrea Modigliani and Paolo Goldoni for their support in data reduction. We also thank Christophe Martayan and Rudy Wij\-nands for helpful discussions as well as Norberto Castro and Grzegorz Pietrzy\'nski for communicating updated photometry. We thank the referee for constructive comments.

\bibliographystyle{mn2e_fix}
\bibliography{scriptie}

\end{document}